# Development of low frequencies, insulating thick diaphragms for power MEMS applications

## *Corresponding author*

Fabien Formosa
Université de Savoie – Laboratoire SYMME, 5 chemin de Bellevue, 74944 Annecy le Vieux, France
Tel : 334 50 09 65 08
Fax : 334 50 09 65 43

## *Authors*

Fabien Formosa, Adrien Badel, Hugues Favrelière

Université de Savoie – Laboratoire SYMME, 5 chemin de Bellevue, 74944 Annecy le Vieux, France

## *Abstract*

Major challenges of micro thermal machines are the thermal insulation and mechanical tolerance in the case of sliding piston. Switching from piston to membrane in microengines can alleviate the latest and lead to planar architectures. However, the thermal isolation would call for very thick structures which are associated to too high resonant frequencies which are detrimental to the engine performances. A thermal and mechanical compromise is to be made. On the contrary, based on fluid structure interaction, using an incompressible fluid contained in a cavity sealed by deformable diaphragm it would be possible to design a thick, low frequency insulating diaphragm. The design involves a simple planar geometry that is easy to manufacture with standard microelectronics methods. An analytical fluid-structure model is proposed and theoretically validated. Experimental structures are realized and tested. The model is in agreement with the experimental results. A dimensionless model is proposed to design hybrid fluid structures for micromachines.

## *Highlights*

- One major challenge of micro thermodynamics machines is to keep high temperature difference.
- Membrane avoids leakage and mechanical friction of piston but is not insulating and lead to too high resonant frequencies.
- We propose an innovative membrane architecture which allows efficient thermal isolation and low resonant frequency.
- We model, design and tested the fluid-structure membranes which are validated.

## *1- Introduction*

To fulfill the demand for high power density microgenerators, the micro engine developments aim at providing devices that deliver electrical power ranging from dozens of milliwatts to a few watts [1-3]. By MEMS approaches, it would be possible to produce these devices which would ultimately be integrated to global units of sensors, actuators and microprocessor as "power plants on a chip". One of the major challenges is the fabrication tolerances of mechanical moving parts of micromachines. Friction losses and leakage through the cylinder-piston gap was underlined as one of the main limitations [2]. Sealing and friction require tolerances that are hardly achieved yet. Moreover, the thermal insulation requirements lead to the use of materials for which processes are under development [3].

Diaphragms are attractive and common structures in MEMS technology mostly used as micropump's components. Switching from sliding piston to flexural diaphragm allows skirting around the classical difficulties associated to the miniaturization of engines. By doing this, sealed, frictionless variable volume chamber can be manufactured.

The application of these structures for power MEMS has been recently studied. Following a new paradigm, phase change micro heat engines (PCMHE) may present the most promising potential mainly because of their simple planar mechanical architecture [4]. Figure 1a presents the generic architecture of these engines. To date, several prototypes were developed. However, these devices have shown poor efficiencies. [5-8].

The operation of phase change micro heat engine is achieved alternately conducting heat in and out of the two-phase working fluid. Starting from flat membranes, the thermodynamic cycle can be described by four processes: a) heat addition, b) expansion, c) heat rejection and d) compression. Depending on the nature of the involved processes, organic Rankine cycle, Carnot or novel thermodynamic cycle could be achieved [7, 8]. Heat dissipation appears again to have the most detrimental effect on the efficiency [4]. The operating dynamics also play a crucial role in the engine optimization. A recent scaling study based on experimental results shows that a low diaphragm resonant frequency increases the performances of a PCMHE [9]. In this latter work, a mass is added to a 10 mm side length and 2.2 µm thick square membrane which allow its resonance frequency to decrease to 100Hz. It is worthy of note that in this case, an insulating polymer material is used to reduce the thermal conduction to the mass and its thermal inertia effect eventually.

As a consequence, the diaphragm technical requirements associated to the PCMHE must exhibit: i) low resonance frequency; ii) high thermal insulation (at least for a one of the two diaphragms).

In addition to the PCMHE, Stirling machines can benefit from diaphragm architectures. The reversible Stirling cycle offers numerous possible applications. A Stirling micro device would typically show millimeters dimensions and few cubic-millimeters swept volume. It could be used for refrigeration or power generation purposes. From its simple architecture (external heat source, no valves, no pump), and alleged high potential performances, it would indeed be well suited as a micro thermal generator.

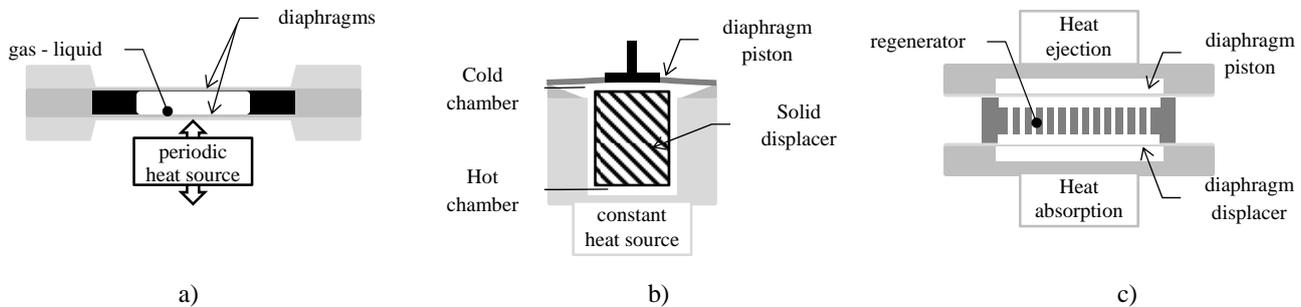

Figure 1: a) typical PCMHE architecture, b) Basics of Harwell TMG Stirling architecture from [10], c) Schematic of Moran Micro Stirling machine from [13]

Few concepts of diaphragm Stirling engine were proposed. The Harwell TMG Stirling engine [10, 11] used a large diameter thick diaphragm made from rubber or polymer. The basics of this engine are presented in Fig. 1b. It benefits from the low stiffness of the diaphragm material to achieve large swept volume. Though, the rubber and polymer material which were used induced important mechanical dissipation. Because of thermal isolation between the cold and hot chambers, the displacer was realized using a classical sliding component. For miniaturization purpose this component may be replaced by a flexural membrane while maintaining high thermal isolation.

A membrane micro-Stirling engine was proposed by Bowman [12]. In a previous work, the authors proposed a model of a membrane Stirling micromachine [13]. Dynamics as well as thermodynamics modeling allow the performances to be predicted. The drastic effect of the diaphragm on the operation of the engine was pointed out. More specifically, based on the specific dynamics free piston Stirling engine, the operating frequency, the strokes and phase angle are related to the membrane's dynamical properties (stiffness, mass). The starting, operation and optimization of the Stirling machine is then strongly coupled to the membrane dynamic design leading to multiple parameters requirements. More recently, for cooling application an alpha Stirling was designed and tested [14, 15] (see Fig. 1c). This Stirling machine presents two 10 mm circular membrane driven by piezoelectric actuators at a phase shift of 90°C. Although, various regenerator structures and materials were evaluated, no experimental operation was demonstrated. Insufficient thermal insulation and large dead volumes are pointed out as the essential cause of failure. Maximizing the swept volume would allow the dead volume impact to be reduced. Finally, it can be inferred that besides the two previous requirements i) and ii), a third requirement iii) is the maximization of the swept volume.

We propose here a new concept of a hybrid fluid diaphragm (HFD) which fulfills the aforementioned needs. It consists in an incompressible fluid confined between two thin membranes. This architecture includes a plain planar geometry that is easy to manufacture with standard microelectronics methods. Figure 2 described its main geometry as well as its geometric parameters. The height of the structure can be tuned to provide sufficient heat isolation. The inertia effect of the fluid allows operation frequencies significantly lower than for a single membrane. Moreover, in addition to the membrane mechanical parameters (geometrical and material properties) the fluid characteristics (height of the cavity, density) can be used for optimization.

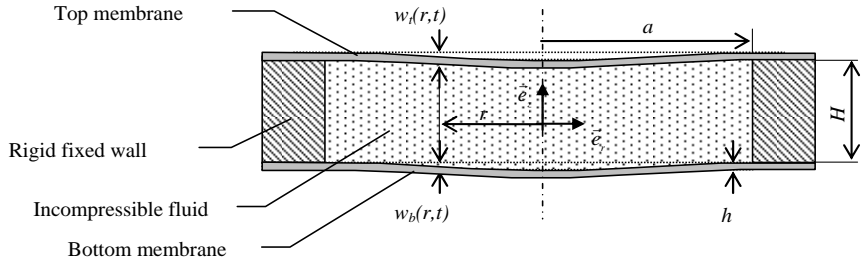

Figure 2: Hybrid fluid diaphragm schematic

Fluid structure interaction problems are extensively studied in engineering. The particular case of two circular plates vibrating in contact with fluid has recently been studied by Jeong [16]. Weak solutions can be found for these problems [17] and finite element [18] or Ritz method [16, 19] can be used for resolution. In the latter, the accuracy of the solution is related to the number of function used for approximation. As an example, the solution is sought using Bessel decomposition in [16]. By doing this, it is possible to determine both the in phase (IP) and out of phase (OP) modes of the problem. However, in the present application, only the first IP mode also called the piston mode is of interest and a reduced simple model can be deduced. In addition, the filling of the container can be performed at a given pressure which induces static deformations of the membranes. For thin membranes this prestress is associated to tension effect and raises the natural frequencies. This can be seen as an additional control parameter for frequency tuning.

In this paper, a simple fluid structure model is proposed. Its accuracy to predict IP modes is demonstrated using FE and literatures models. Then, the model is compared to experimental results for which the static prestress effects due to the fluid filling process have to be taken into account. Finally, a dimensionless formulation is presented. It allows optimization with respect to membrane micro-machines technical requirements.

## *2- Fundamental equations and method of resolution*

The modeling of the fluid structure problem is based on potential fluid flow theory. The filling fluid is assumed to be an incompressible and inviscid liquid. Elastic cylindrical diaphragms are considered here and prestress from static deformations of the diaphragms are taken into account. Moreover, we consider large amplitude vibrations of the mechanical structures. The analysis is restricted to axially symmetric motions.

### *2.1- Fluid equations*
With the liquid motion assumed to be irrotational, the governing equation for the liquid results in:

$$\nabla^2 \phi(r,z,t) = 0 \quad \forall r \in [0,a], \forall z \in \left[-\frac{H}{2}, \frac{H}{2}\right]$$

(1)

In which $\Phi(r, z, t)$ is the velocity potential such as the fluid velocity is defined as: $\vec{U} = \vec{\nabla}\phi(r,z,t)$. $\vec{\nabla}$ and $\nabla^2$ are the cylindrical Gradient and Laplace operators respectively.

At the side wall, the velocity satisfies:

$$\frac{\partial \phi}{\partial r} = 0 \text{ for } r = a, \forall z \in [-H/2, H/2], \forall t \in [0,T]$$

(2)

At the bottom and top, the fluid velocity matches the diaphragm's one:

$$-\frac{\partial \phi}{\partial z}(r,-\frac{H}{2},t) = \frac{\partial w_b}{\partial t}(r,t) \quad \forall r \in [0,a], \forall t \in [0,T]$$

$$-\frac{\partial \phi}{\partial z}(r,\frac{H}{2},t) = \frac{\partial w_t}{\partial t}(r,t) \quad \forall r \in [0,a], \forall t \in [0,T]$$

(3)

In which $\frac{\partial w_p}{\partial t}$ are the bottom and top diaphragms velocity with $p = b$ and $t$ for the bottom and top diaphragms defined in Fig. 2.

### *2.2- Diaphragm equations*
The main assumptions used for the modeling of the membrane are the following:

- Membranes are submitted to mechanical solicitations such as their strains are axisymmetric only.
- The midplane displacements are in the $z$ direction and denoted $w_b(r,t)$ for the bottom and $w_t(r,t)$ for the top diaphragms.
- Plane stresses are considered (shear stresses are neglected).
- Rotational kinetic energy is neglected.
- Top and bottom membranes are supposed to be identical.

The dynamic equilibrium equation for a clamped circular diaphragm is then:

$$D\left(\frac{\partial^2 w_p}{\partial r^2} + \frac{1}{r}\frac{\partial w_p}{\partial r}\right)\left(\frac{\partial^2 w_p}{\partial r^2} + \frac{1}{r}\frac{\partial w_p}{\partial r}\right) = \frac{\partial^2 w_p}{\partial t^2} \tag{4}$$

In which, $D = E h^3 / 12 (1-\nu^2)$, $h$ is the diaphragm thickness, $\rho$, $E$ and $\nu$ its density, Young modulus and Poisson coefficients respectively.

The diaphragm is assumed to be clamped at the wall, though:

$$w_p(a,t) = 0 \quad \forall t \in [0,T]$$
$$\frac{\partial w_p}{\partial r}(a,t) = 0 \quad \forall t \in [0,T] \tag{5}$$

### 2.3- Method of solution for the fluid problem

A Ritz method is used to find approximated solution to the coupled fluid-structure problem. For the fluid part, the approximated solution will be sought as an extended solution of Eq. (1) and (2). The general form of solutions for Eq. (1) is:

$$\phi_g(r,z,t) = \sum_{j=1}^{N}\left(C_j(t)\sinh(k_j z) + D_j(t)\cosh(k_j z)\right) J_0(k_j r) \tag{6}$$

In which $N$ is the number of retained solution of the transcendental equation associated to Eq. (2) and $J_0(x)$ is the Bessel first kind function. Taking into account the velocity symmetry condition for the IP modes, the solutions can be simplified as:

$$\phi(r,z,t) = \sum_{j=1}^{N} C_j(t)\sinh(k_j z) J_0(k_j r) \tag{7}$$

Approximate solutions $\bar{\phi}(r,z,t)$ for the fluid problem are sought using a solution of Eq. (2) such as:

$$\bar{\phi}(r,z,t) = C_1(t) z + \sum_{j=1}^{N} C_{j+1}(t)\sinh(k_j z) J_0(k_j r) \tag{8}$$

By doing this, an enriched function which enhances the approximation while satisfying the side wall boundary condition (Eq. (2)) is obtained.

For the diaphragm, the linear bending eigenmodes of the diaphragm $\bar{w}_i(r)$ satisfying Eq. (4) and (5) can be expressed. Therefore, we set:

$$w_p(r,t) = \sum_{i=1}^{M} p_i(t)[\bar{w}_i(r)] = \sum_{i=1}^{M} p_i(t)\left[J_0(\beta_i r) - \frac{J_0(\beta_i a)}{I_0(\beta_i a)} I_0(\beta_i r)\right] \tag{9}$$

In which $p_i$ is the generalized coordinate, $\omega$ is the pulsation and $I_0(x)$ is the modified Bessel function. $k_j$ satisfies Eq. (2). $\beta_i$ is the solutions of the transcendental equation associated to Eq. (5) such as $\beta_i^4 = \omega_i^2 \rho e_d / D$. $M$ is the number of modes to be selected for the approximation of the deflection.

For developing the weak form of Eq. (1) and (3), we use the function $\tilde{\phi}(r,z) = \tilde{C}_1 z + \sum_{j=1}^{N}\tilde{C}_{j+1}\sinh(k_j z) J_0(k_j r)$ and integrate over the fluid domain. Using boundary conditions we obtain:

$$-\int_{-H/2}^{H/2}\int_0^a \left(\frac{\partial \bar{\phi}}{\partial r}\frac{\partial \tilde{\phi}}{\partial r} + \frac{\partial \bar{\phi}}{\partial z}\frac{\partial \tilde{\phi}}{\partial z}\right) r\, dr\, dz + 2\int_0^a \left(\sum_{i=1}^{M}\dot{p}_i(t)\bar{w}_i(r)\right)\tilde{\phi}(r,H/2) r\, dr = 0\ \forall \tilde{C}_i, \forall t \in [0,T], i = 1..N \tag{10}$$

The coefficients $C_j$ are then found from Eq. (10). In the case of a 7 mm radius membrane, Fig. 3 presents the shapes of the top diaphragm for the first three IP modes ($\bar{w}_i(r)$ $i$=1..3) and the corresponding values for the boundary condition of the fluid problem along the radius ($-\partial\bar{\phi}_i/\partial z$ $i$=1..3 see Eq. (3)). Note that the amplitudes of the mode shapes rely on their direct evaluation without normalization. From the discrepancies shown in Fig. 3, it can be inferred that the simple approximation of the velocity potential is good enough to satisfy the condition especially for the first and second IP modes. Hence, the dynamical fluid-structure coupling is expected to be well accounted for.

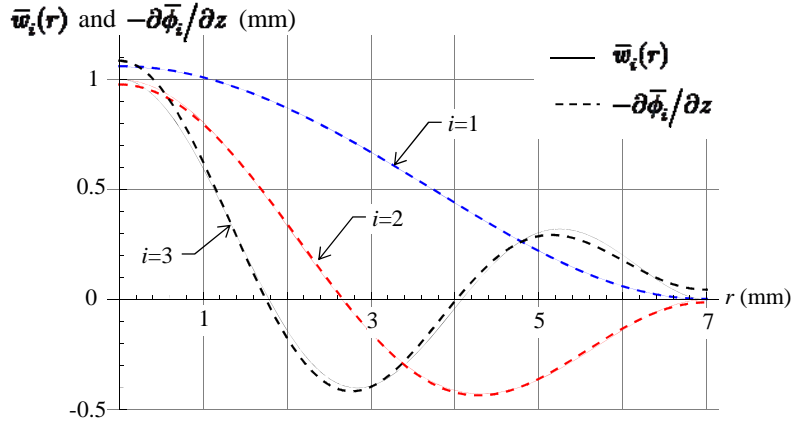

Figure 3: Shapes for the 3 first IP eigenmodes $\bar{w}_i(r)$ (plain lines) and corresponding approximate solution for the fluid problem $-\partial \bar{\phi}_i / \partial z$ (dashed lines)

*2.4- Dynamics of diaphragms*
To the extent that the harmonic motion is supposed, strain and kinetic energy for the mechanical problem can be obtained using a basis of linear eigenmodes (Eq. (9)) which compose the spectral basis to express the transverse motion. In the case of large deflection, nonlinear geometric behavior leads to a tensor of the fourth order associated to in plane induced tension in addition to the classical stiffness and mass ones. Initial prestress is also taken into account. Based on the previous results, the fluid motion can be expressed as a function of the amplitude of the first modes of the top and bottom diaphragms. Therefore, the fluid interaction with the diaphragm can be added to the problem.

From the application of the Hamilton's principle for each diaphragm, sets of algebraic nonlinear equations can be derived. Provided that the spectral basis is limited to the first mode, this approach gives an analytical model of the nonlinear fluid coupled dynamical behavior. The equilibrium equations are obtained using the Hamilton's principle:

$$d \int_0^T \left( E_c - E_{dL} - E_{dNL} - W_f - W_s \right) dt = 0 \tag{11}$$

Where, $E_c$ is the kinetic energy, $E_{dL}$ the linear strain energy, $E_{dNL}$ the nonlinear strain energy given in [20] and $W_f$ is the work of the fluid forces related to the pressure on the considered diaphragm. $W_s$ is the work associated to an initial pre-stress. *0* and *T* are times such as no deflection of the diaphragm occurs. $E_c$, $E_{dL}$, $E_{dNL}$, $W_f$ and $W_s$ can be expressed as follow:

$$E_c = \tfrac{1}{2} \dot{p}_i \dot{p}_j 2\pi \rho h \int_0^a \bar{w}_i \bar{w}_j r\, dr = \tfrac{1}{2} p_i p_j m_{ij} \tag{12}$$

$$E_{dL} = \tfrac{1}{2} p_i p_j 2\pi D \int_0^a \frac{\partial^2 \bar{w}_i}{\partial r^2} \frac{\partial^2 \bar{w}_j}{\partial r^2} + \frac{1}{r^2} \frac{\partial \bar{w}_i}{\partial r} \frac{\partial \bar{w}_j}{\partial r} r\, dr = \tfrac{1}{2} p_i p_j k_{ij} \tag{13}$$

$$E_{dNL} = \tfrac{1}{2} p_i p_j p_k p_l 2\pi \frac{3D}{h^2} \int_0^a \frac{\partial w_i}{\partial r} \frac{\partial w_j}{\partial r} \frac{\partial w_k}{\partial r} \frac{\partial w_l}{\partial r} r\, dr = \tfrac{1}{2} p_i p_j p_k p_l b_{ijkl} \tag{14}$$

In which $m_{ij}$, $k_{ij}$ are the mass and stiffness matrix, $b_{ijkl}$ the nonlinear stiffness matrix. In the considered application, the work of the fluid dynamic pressure forces is:

$$W_f = p_i 2\pi \rho_f \int_0^a \left. \frac{\partial \bar{\phi}}{\partial t} \right|_{z=z_p} \bar{w}_i(r,t)\, r\, dr = p_i f_i \tag{15}$$

$$W_s = p_i p_j 2\pi h \int_0^a \sigma_{rs} \frac{1}{2} \frac{\partial \overline{w}_i}{\partial r} \frac{\partial \overline{w}_j}{\partial r} r\, dr = p_i p_j n_{ij} \tag{16}$$

In which $f_i$ is the pressure coefficient, $\sigma_{rs}$ is the prestress of the diaphragm and $n_{ij}$ the related matrix. Replacing $E_c$, $E_{dL}$, $E_{dNL}$, $W_f$ and $W_s$ by their discretized expressions into Eq. (11) and taking into account the symmetry of the IP modes, $M$ sets of equations can be obtained as:

$$\left(m_{ij} + 2\frac{\partial W_{fi}}{\partial \ddot{p}_i}\right)\ddot{p}_i + \left(k_{ij} + n_{ij}\right)p_i + 2b_{ijkl}p_i p_j p_k + 2\ddot{p}_j \frac{\partial W_{fi}}{\partial \ddot{p}_j} = 0 \tag{17}$$

In the case of one single mode and one retained solution of Eq. (2) $k_{11}$, assuming harmonic motion, we obtain:

$$\ddot{p}_1\left(m_{11} + \frac{\partial W_{f1}}{\partial \ddot{p}_1}\right) + p_1\left(k_{11} + n_{11}\right) + 2 p_1^3 b_{1111} = 0 \tag{18}$$

Thus, assuming small deflection, the first IP mode frequency of the HFD is:

$$f = \frac{1}{2\pi}\sqrt{\left(k_{11} + n_{11}\right) \bigg/ \left(m_{11} + \frac{\partial W_f}{\partial \ddot{p}_1}\right)} \tag{19}$$

Eq. (19) clearly demonstrates the inertia effect of the fluid ($\partial W_f / \partial \ddot{p}_1$) whereas the prestress ($n_{11}$) increases the effective stiffness.

## *3- Results and discussions*

### *3.1- Model theoretical validation*
The proposed method is validated comparing the results to an ANSYS® FE model as well as to the results from Jeong's study [16]. Table 1 summarizes the test case definition as well as material parameters. Note that no prestress is taken into account and linear behavior is considered here.

| Diaphragm |  |  |  |  | Fluid |  |
|---|---|---|---|---|---|---|
| Radius | Thickness | Young modulus | Poisson ratio | Density | Height | Density |
| $a = 120$ mm | $h = 2$ mm | $E = 69$ GPa | $\nu = 0.3$ | $\rho = 2700$ kgm$^{-3}$ | $H = 40$ mm | $\rho_f = 1000$ kgm$^{-3}$ |

Table 1: Test case parameters

Table 2 presents results for the first three IP modes for different values of $N$ and $M$.

| | FE model | Jeong [16] | | **M = 1**<br>**N = 1** | M=2<br>N=1 | **M = 2**<br>**N = 2** | M=3<br>N=1 | M=3<br>N=2 | **M = 3**<br>**N = 3** |
|---|---|---|---|---|---|---|---|---|---|
| IP mode 1 | | | 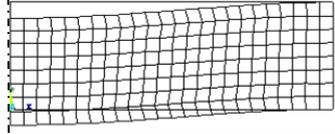 | | | | | | |
| Frequency (Hz) | 162 | 162.8 | | **162.9** | 162.9 | **162.85** | 162.9 | 162.85 | **162.85** |
| Discrepancy (%) | | -0.49 | | **0.55** | 0.55 | **0.52** | 0.55 | 0.52 | **0.52** |
| IP mode 2 | | | 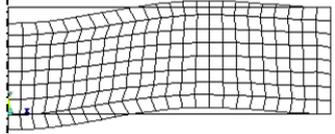 | | | | | | |
| Frequency (Hz) | 666.9 | 677.5 | | | 874.12 | **678.65** | 860.92 | 678.64 | **677.60** |
| Discrepancy (%) | | 1.59 | | | 31.07 | **1.76** | 29.10 | 1.76 | **1.65** |
| IP mode 3 | | | 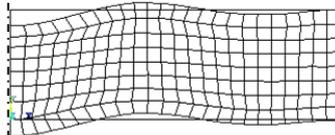 | | | | | | |

| Frequency (Hz) | 1602 | 1642.8 | | | | | 2749.39 | 2127.4 | **1648.51** |
|---|---|---|---|---|---|---|---|---|---|
| Discrepancy (%) | | 2.54 | | | | | 71.62 | 32.79 | **2.90** |

Table 2: Comparison between FE, Jeong's study and proposed approach analytical results

Results show that the fluid problem approximation basis ($N$) has to match the number of diaphragm modes that are used for the mechanical behavior approximation ($M$). Indeed, the chosen form of the test function $\tilde{\phi}(r,z,t)$ has to be close enough to the modes shapes of the diaphragm to satisfy the boundary Eq. (3) whereas Eq. (2) is ensured by the chosen $k_i$ values.
In these cases, the maximal discrepancy is 2.9 % for the third IP mode. Using larger basis for the test functions would allow the accuracy to be enhanced.

Additionally, these results show that as long as the first IP mode only is sought, the simplified model using $M=1$ and $N=1$ gives an accurate evaluation of the HFD frequency.

*3.2- Model experimental validation*
Experimental validation has been performed. Prototypes of HFD have been made from 25 µm thickness Kapton®. The Young modulus, Poisson's ratio and density values for the diaphragms are $E = 5.5$ GPa, $v = 0.34$ and $\rho = 1420$ kg m$^{-3}$. The Kapton® film is glued on a rigid aluminum case and the interspace cavity is filled with water and sealed. The aluminum frame is 9.2 mm height and cavities diameters range from 7 to 9 mm. The HFD can be clearly seen on the picture of the experimental mockup given in Fig. 4.
The HFD is excited by an electrodynamical shaker driven with an increasing frequency sweep. Using an accelerometer and a dedicated control unit, the shaker is driven in closed-loop to ensure constant acceleration amplitude. The displacement of the top membrane is measured using an embedded laser optical sensor Micro-Epsilon LD1607-0.5 which sensitivity is 40V/mm and resolution is 0.1 µm.

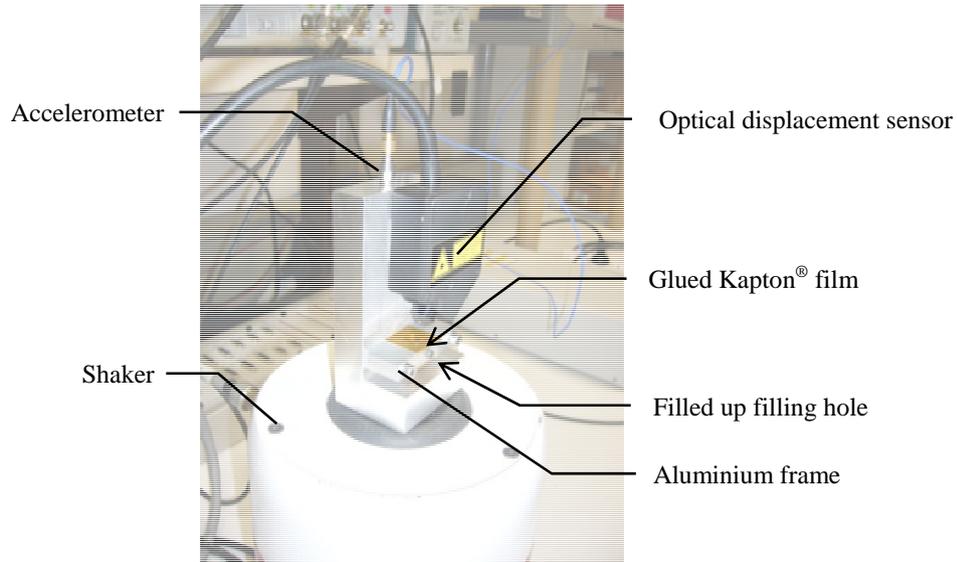

Figure 4: Experimental setup for model validation

To prevent any air bubble to be trapped into the cavity which would lead to a dramatically degraded behavior of the system (a compressible volume of gas would act as an additional stiffness leading to out of phase motions of the top and bottom membranes), a filling fluid volume larger than the cavity is injected and a screw seals it. Therefore, the water pressure increases as the volume diminishes causing the diaphragm to bulge outward. The dynamic motions of the membranes are then small amplitude oscillations around their deformed shape. Because we plan to optimize the fluid filling process to obtain flat membranes HFD, we choose here not to develop a devoted model which takes into account the static large deflection. Nevertheless, induced stresses are included in the model as the prestress $n_{11}$ term.

Figure 5 shows the static measurements performed for 7, 8 and 9 mm diaphragms. For this study, a surface measuring machine based on extended field confocal microscopy is used, equipped with a confocal chromatic optical probe with 300 µm depth of field, and 0.06 µm axial accuracy.

It is worthy of note that due to the assembling process, the devices are not necessarily symmetrical. Some experimental flaws can be seen in Fig. 5. If the bonding of the Kapton® is not strong enough, fluid can flows between the film and the aluminum frame surface. This is the case for the 8 mm bottom diaphragm (left picture of Fig. 5b). On the contrary if the glue happens to go over the edge of the hole, the membrane is retained and the static deformation when submitted to pressure presents complex shape as it is the case for the 9 mm top diaphragm (right picture of Fig. 5c).

The important differences between the real and the assumed initial ideal shape lead to evaluate an effective radius to compare the model to experimental results. We assume that the shape of the deflected surface $w_s(r)$ is expressed as Eq. (20), which correspond to the theoretical shape of a diaphragm under a static load:

$$w_s(r) = w_{s0}\left(1 - r^2/a^2\right)^2 \tag{20}$$

In which $w_{s0}$ is the maximal static deflection and $a$ the radius of the diaphragm.

By doing this, the theoretical volume for a given maximal deflection can be expressed as $V_{th} = \pi\, w_{s0}\, a^2/3$. The volume of the measured bump $V_{exp}$ is numerically evaluated from the static measurements and an effective radius $a_{eff}$ is found such as the two volumes are the same. Table 3 gives the values of the effective radius for the diaphragms.

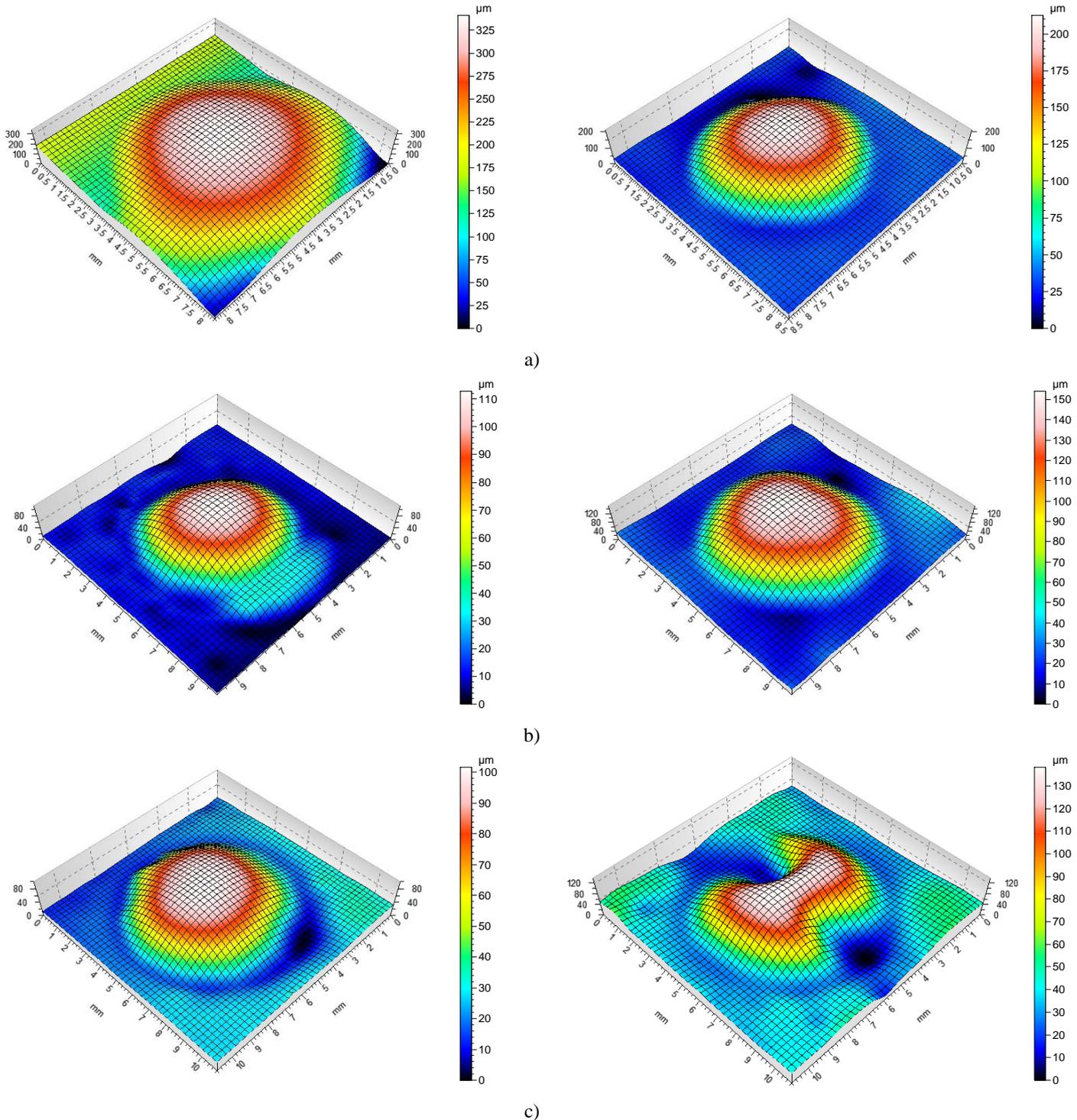

a)

b)

c)

Figure 5: a) Initial static sag measurements of top and bottom 7 mm HFD, b) 8 mm HFD, c) 9 mm HFD

From the static deflection of the diaphragm, it is possible to evaluate the stresses of the structures. In addition to the deflection, we assume that the radial displacement $u_s(r)$ is given by [21]:

$$u_s(r) = r(a - r)(A_1 + A_2 r) \quad (21)$$

In which, $A_1 = 1.185\, w_{s0}^2/a^3$ and $A_2 = -1.75\, w_{s0}^2/a^4$.

Finally, the radial stress as a function of the measured central displacement is:

$$\sigma_{rs} = \frac{E}{1-\nu^2}(\varepsilon_{rs} + \nu\, \varepsilon_{\theta s}) = \frac{E}{1-\nu^2}\left( \frac{\partial u_s}{\partial r} + \frac{1}{2}\left(\frac{\partial w_s}{\partial r}\right)^2 + \nu \frac{u_s}{r} \right) \quad (22)$$

From Eq. (22), the prestress effect can be evaluated and the membranes natural frequencies (without fluid interaction) updated (see Table 3).

|  | Bottom diaphragm | | | Top diaphragm | | | First IPM of the HFG | | |
|---|---|---|---|---|---|---|---|---|---|
| Diaphragm | diameter (mm) | Max. deflection (µm) | Natural frequency (kHz) | diameter (mm) | Max deflection (µm) | Natural frequency (kHz) | exp. frequency (Hz) | theo. frequency (Hz) | Discrepancy (%) |
| 7 mm | 5.83 | 341 | 20.5 | 5.82 | 212 | 13.02 | 1650 | 1736 | 5.2 |
| 8 mm | 6.31 | 113 | 6.25 | 7.32 | 155 | 6.47 | 1000 | 702 | 29 |
| 9 mm | 7.37 | 102 | 4.22 | 8.00 | 138 | 4.63 | 421 | 495.3 | 17.6 |

Table 3: results

Figure 6 presents the experimental dynamic responses for the 3 tested HFD. The experimental identified resonance frequencies are clearly visible. Their values are reported in Table 3. The fluid inertia effect is very important and allows the natural frequencies to dramatically lessen by a factor of about 10. As expected, the frequencies decreases when the HFD diameters is larger.

Low frequency behaviors are below the sensor's resolution for the 7 and 8 mm HFD which does not affect the natural frequency evaluations. Though high quality factor of the HFD was not sought, these tests show pretty sharp resonance peaks. With optimized assembling process, the quality factor is expected to increase.

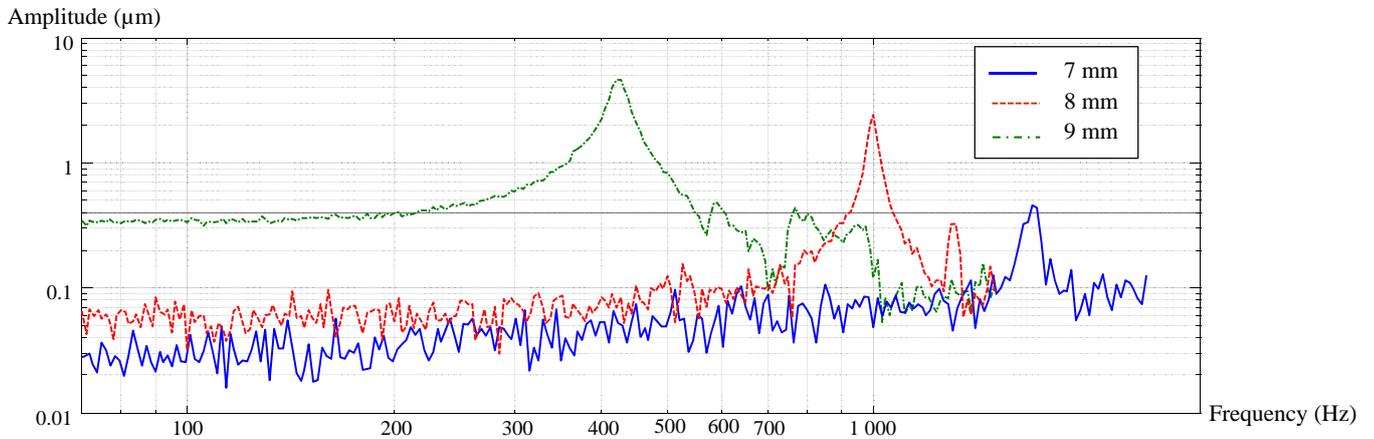

Figure 6: Experimental measurement of membrane displacement with respect to frequency

The mean value of the discrepancies between the model and the experimental results is 17.3 %. It is the author's feeling that the main cause for the error is associated to the technical difficulties for gluing the Kapton® and during the filling process (compare Figures 5a-5b and Fig. 5c). Nevertheless, in spite of the geometrical differences and model assumption, a pretty good agreement is found and the model can then be used as a design tool.

This first experimental realization of HFD attempted to evaluate the model capacity to predict the resonant frequency. The damping ratio or quality factor optimization has not been sought for in this experimental device. Though, based on the results, a prior estimation of the HFD damping or quality factor can be made. The quality factors range from 15 *(evaluation 9 mm)* to 38 *(evaluation 8 mm)*. Again, the gluing process as well as the polymer material selection will be optimized to minimize the damping. Further investigations would allow distinguishing the effect of the fluid from the membrane on the damping.

## *4- Design and optimization*

A simplified model ($M = 1$ and $N = 1$) to predict the first IP mode has been validated. Based on this model, it is possible to analytically express the HFD frequency and perform easily multi-criteria optimization for micro-engine applications. We choose to use a dimensionless formulation. By doing this, the model can be used for centimeter or millimeter scales HFD. For the sake of clarity, no prestress is considered hereafter, though its effect could be easily taken into account.

### *4.1- Dimensionless expression*

The dimensionless variables are defined such as: $r = a\, r^*$; $z = H\, z^*$; $w_p(r,t) = h\, w_p^*(r^*, t^*)$ and $\phi(r,z) = a\, H\, \phi^*(r^*,z^*)$.
By doing this, Eq. (10) can be rewritten as:

$$-a^2 H^3 \int_0^1 \int_{-1/2}^{1/2} \left(\frac{\partial \overline{\phi}^*}{\partial r^*}\right)^2 r^* \, dr^* \, dz^* - a^4 H \int_0^1 \int_{-1/2}^{1/2} \left(\frac{\partial \overline{\phi}^*}{\partial z^*}\right)^2 r^* \, dr^* \, dz^* + 2\left(ha^3 H\right) \dot{p}_1 \int_0^1 w_b^*(r) \overline{\phi}(r, -\tfrac{1}{2}) r^* \, dr^* = 0 \, \forall C_i, i = 1..N \quad (23)$$

Evaluating the integrals in Eq. (23), the coefficients $C_1$ and $C_2$ are given by:

$$C_1 = -0.3289 \dot{p}_1 \frac{h}{a} \quad C_2 = -0.4132 \dot{p}_1 \frac{ha}{2.9878 H^2 + 4.1787 a^2} \quad (24)$$

The nonlinear dynamic equilibrium is then:

$$D \left( k_{11}^* + 2 b_{1111}^* \left( p_1^* \right)^2 \right) p_1^* - a^4 \omega^2 \left( \rho h m_{11}^* + \rho_f H \left( 0.2983 - \frac{0.0918 H^2}{0.715 H^2 + a^2} \right) \right) p_1^* = 0 \quad (25)$$

In which $m_{11}^* = 2\pi \int_0^1 w_b^* w_b^* r^* \, dr^* = 0.6402$, $k_{11}^* = 2\pi \int_0^1 \left[ \left(\frac{\partial^2 w_b^*}{\partial r^{*2}}\right)^2 + \frac{1}{r^{*2}} \left(\frac{\partial w_b^*}{\partial r^*}\right)^2 \right] r^2 \, dr^2 = 66.81$, $b_{1111}^* = 3 \int_0^1 \left(\frac{\partial w_b^*}{\partial r}\right)^4 r^* \, dr^* = 26.97$.

From Eq. (25), it is now possible to obtain an analytical expression for the HFD linear natural frequency:

$$\omega_1 = \frac{h}{a^2} \sqrt{\frac{E}{12(1-\nu^2)\rho}} \sqrt{66.81 \Big/ \left( 0.6402 + \frac{\rho_f H}{\rho h} \left( 0.2983 - \frac{0.0918 H^2}{0.715 H^2 + a^2} \right) \right)} \quad (26)$$

As a first example, we illustrate how the design of the hybrid diaphragm can be used in the design of a PCMHE. Starting from some of the characteristics of the P3 micro engine [10], we consider a 2.2 µm thick Silicon membrane whose diameter range from 8 to 10 mm. Comparing the resonant frequency of a single membrane $\left( h/a^2 \sqrt{E/12(1-\nu^2)\rho} \sqrt{66.81/0.6402} \right)$, the added inertia of the fluid allows it to be reduced as stated by Eq. (26). By lowering the resonant frequency, the performances of the engine can be improved. Figure 7 shows the resonance frequencies of the membrane and the HFD with respect to the height of the fluid cavity $H$ for two radii 5 and 4 mm.

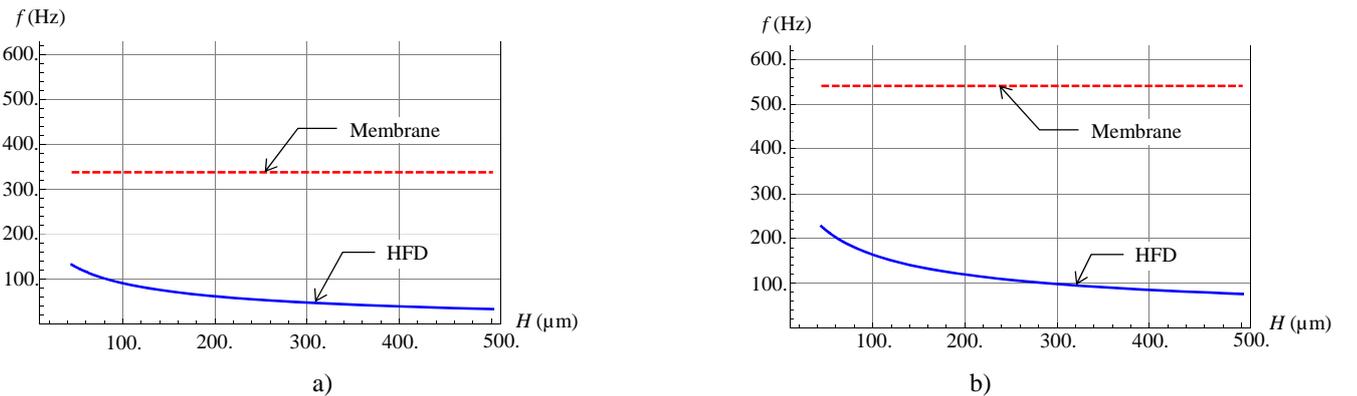

Figure 7: Resonance frequencies of diaphragm and HFD as a function of *H* for a) 5 mm radius, and b) 4 mm radius

With moderate heights for the fluid cavity it is possible to match the low resonant frequency requirements of the PCMHE. For radius of 5 mm, the HFD frequency is below 100 Hz with 100 microns height of fluid whereas 300 microns are needed for 4 mm radius. It is worthy of note that the HFD approach fits the fabrication process used for the PCMHE with no need of additional mass to obtain low natural frequencies[6, 9].

*4.2- Design and optimization*
We proposed hereafter an illustration of a design strategy for a microengine using HFD. It is assumed that the design of micro engine diaphragms can be defined by three requirements:
  i.   A thermal resistance $R_{th}$.
  ii.  A target linear operating frequency $f$ or frequency range $[f_{min}, f_{max}]$
  iii. A swept volume $V_{sw}$.

Considering again the architecture and requirements of the previously described PCMHE, a HFD could be effectively used to replace the upper membrane of the PCMHE described in Fig. 1a. By doing this, the dynamic is conveniently set by the upper HFD with no need of additional mass. Moreover, the thermal insulation is controlled in the upper part whereas the heat transfer surface is limited to the thin bottom membrane. The operation and performances of such a thermally oscillation engine also rely on the thermal mass of its different component. Though the HFD would represent a thermal capacity higher than the thin membrane, it would be presumably comparable to the additional mass used for optimization in [9]. Though, the proposed strategy does not focus on this characteristic, the thermal capacity could be set as an additional requirement and fluid properties (notably density and thermal capacity) other design parameters.
In the case of the Stirling engine architecture described in Fig. 1b, the gas leakage through the displacer/casing clearance induces reheat losses due to the flow of cold gas into the hot expansion chamber which impacts the efficiency of the engine. A HFD can be used to replace the solid displacer to suppress the pertaining leakage and friction. In this case the thermal resistance of the HFD has to be tuned to keep the requisite temperature difference between the hot and cold chambers. Moreover, the thermal mass of the HFD is not a critical characteristic for the steady state behavior of the Stirling engine.
Based on this discussion, a representative design test case is proposed as an attempt to demonstrate the relationship between the design parameters: for a given material and thickness of the diaphragm (2.5 µm thick Silicon) and assuming a given requisite thermal resistance, the design of a HFD is sought to match an operating frequency and swept volume.

For a given conductive heat transfer, the thermal resistance through the HFD is governed by the $H/a^2$ ratio. As a result, the radius of the HFD can be defined as a function of the height and the evolution of the frequency and swept volume can be expressed as a function of the height $H$ of the HFD. For the sake of clarity, we focus here on the height as a design parameter, the thickness, material of the membrane as well as the fluid properties can be also used as design parameters though.

*Evolution of the operating frequency with respect to HFD height for a given thermal resistance:*
Fig. 8a presents the evolution of the HFD natural frequency. For comparison purpose, the evolution of the frequency of a single membrane with thickness $H$, is represented as well. The frequency is drastically reduced by the inertia of the fluid. Because of the constant thermal resistance requirement, the radius is modified with respect to $H$ and is plotted in Fig. 8b. For a constant thermal resistance, the radius of the HFD is larger as the height increases. As a result, the natural frequency of the single membrane decreases as the radius increases. This allows the swept volume to be increased accordingly as will be described hereafter.

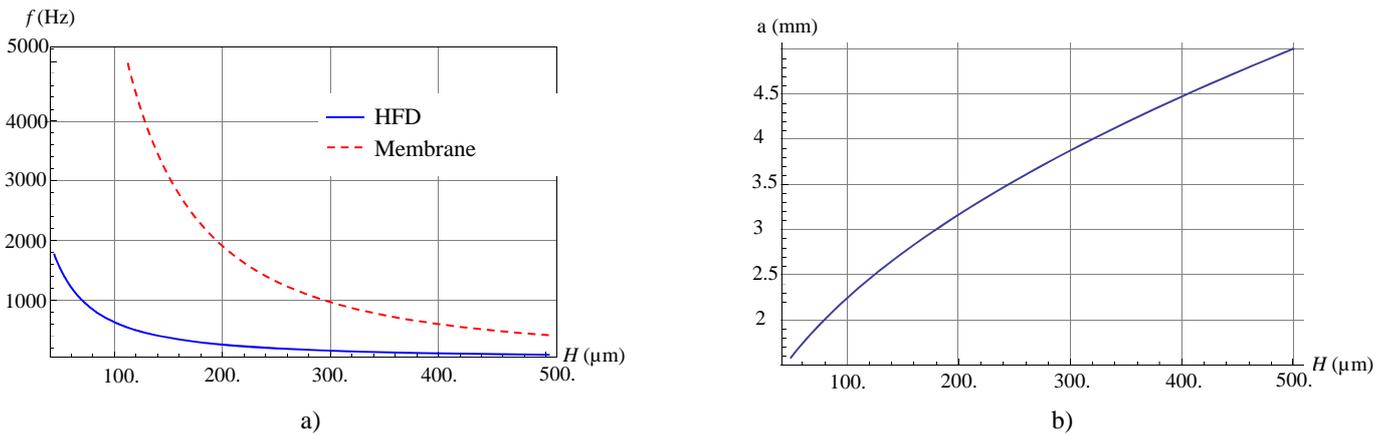

Figure 8: Resonance frequencies of HFD and single membrane as a function of $H$ with constant thermal resistance b) Evolution of the radius of the HFD with respect to $H$ with constant thermal resistance.

*Evolution of the swept volume with respect to HFD height for a given thermal resistance:*
The swept volume is proportional to the surface of the diaphragm multiply by the stroke as long as a linear membrane behavior is assumed. However, the maximal stroke of the diaphragm is limited by the tension effect associated to large displacements especially for resonant micro engine. The evaluation of the ratio between the first two stiffness terms of Eq. (25), $k_{11}$ and $2b^*_{1111}p_1^{*2}$ is relevant to assess the linearity of the stroke. Fig. 9 plots the stiffness ratio $2b^*_{1111}p_1^{*2} / k_{11}$ as a function of the dimensionless static central deflection $w_{s0}/h$.

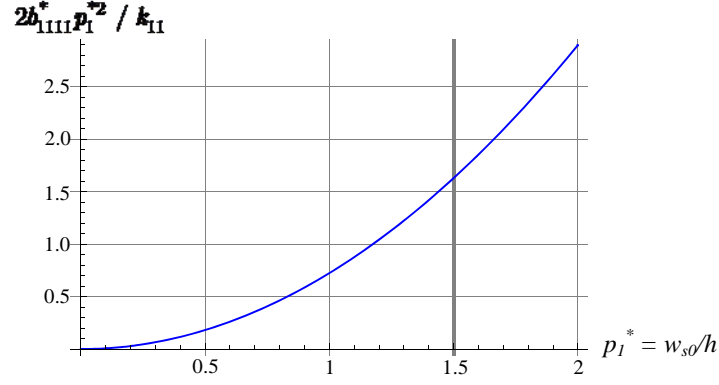

Figure 9: Evolution of the $2b_{1111}^{*} p_1^{*2} / k_{11}$ stiffness ratio with respect to the dimensionless deflection.

It appears that the actual stiffness is more than twice its linear value as soon as the deflection equals about 1.5 the diaphragm thickness. Therefore, for thin membrane and large deflection, the nonlinear tension effects have to be taken into account which is consistent with the literature results [4]. As a consequence, we assume that the order of magnitude of the swept volume can be evaluated for half the critical value of 1.5 $h$. It is expressed as:

$$V_{swept} = 1.5\,h\ 2\pi\,a^2 \int_0^1 w_b^* r^* dr^* \approx 5.1\,h a^2 \qquad (27)$$

For a given constant thermal resistance, the swept volumes as well as the HFD frequency evolutions with respect to $H$ are represented for three different diaphragm thicknesses: 2.2, 4 and 5 µm in Fig. 10. As stated by Eq. (27) the swept volume is proportional to the membrane thickness $h$. As the height of the HFD increases, so is the radius keeping a constant thermal resistance (see Fig. 8b) and the swept volume is larger. It can be seen from Fig. 10 that the HFD natural frequency is not proportional to $h$ because the inertia effect of the fluid diminishes as $h$ increases as stated by Eq. (26). Based on Fig. 10, design areas can be set and prior selection of the basic parameters of the HFD easily performed.

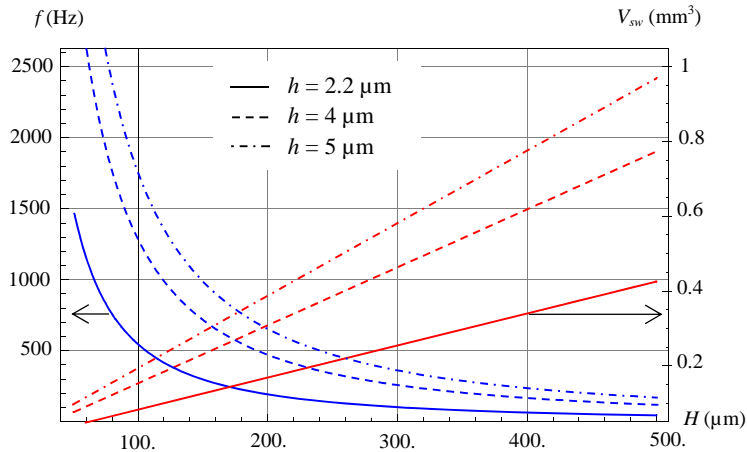

Figure 10: Evolution of the swept volume and HFD frequency as a function of $H$ with constant thermal resistance for various membrane thicknesses.

## 5- Conclusion

A HFD concept very suitable for micro fabrications techniques and which can be used for membrane micro machines has been proposed. Based on fluid-structure interaction, a comprehensive model was proposed and validated. It is able to predict the important frequency characteristics for the first IP or piston mode relevant in the micro engine background. Although the model is developed for cylindrical membranes, the HFD concept can be easily extended to square geometries. Experimental prototypes have been obtained and tested. Even if the filling and bonding process need to be optimized, first results are in accordance with the model predictions. Finally, a dimensionless formulation of the model appears to be a handy tool for HFD design and demonstrates the important benefits of this approach, in the field of thermal micro-engine especially.

## *References*

[1] A. Carlos Fernandez-Pello, "Micro-Power Generation Using Combustion: Issues and Approaches", Twenty-Ninth International Symposium on Combustion, July 21-26 2002, Sapporo, Japan
[2] S.K. Chou, W.M. Yang, K.J. Chua, J. Li, K.L. Zhang, "Development of micro power generators – A review", Applied Energy 88 (2011), pp. 1–16
[3] D. C. Walther, J. Ahn, "Advances and challenges in the development of power-generation systems at small scales", Progress in Energy and Combustion Science 37 (2011), pp. 583-610
[4] L.W. Weiss, "Power production from phase change in MEMS and micro devices, a review", International Journal of Thermal Sciences 50 (2011), pp. 639-647
[5] L. W. Weiss, J.H. Cho, K.E. McNeil, C.D. Richards, D.F. Bahr and R.F. Richards, "Characterization of a dynamic micro heat engine with integrated thermal switch", J. Micromech. Microeng. 16 (2006) S262–S269
[6] S. Whalen, M. Thompson, D.F. Bahr, C.D. Richards, R.F. Richards, "Design, fabrication and testing of the P3 micro heat engine", Sensors and Actuators A 3649 (2003), pp.1–9
[7] J H Cho, LW Weiss, C D Richards, D F Bahr and R F Richards, "Power production by a dynamic micro heat engine with an integrated thermal switch", J. Micromech. Microeng. 17 (2007), pp. S217–S223
[8] T Huesgen, J Ruhhammer, G Biancuzzi and P. Woias, "Detailed study of a micro heat engine for thermal energy harvesting", J. Micromech. Microeng. 20 (2010) 104004 (9pp)
[9] H. Bardaweel, B. S. Preetham, R. Richards, C. Richards, M. Anderson, "MEMS-based resonant heat engine: scaling analysis", Microsyst Technol 17 (2011), pp.1251–1261
[10] E.H. Cooke-Yarborough, "Small Stirling-cycle power sources in marine applications", Oceans'80, 8-10 Sept. 1980, Seattle, USA, pp. 457-462
[11] R. Boukhanouf,R., S.B. Riffat, and R. Shuttleworth, "Diaphragm Stirling engine design", 2$^{nd}$ International Heat Powered Cycles Conference: HPC 2001, Paris, France.
[12] L. Bowman, "Microminiature Stirling cycle cryocoolers and engine", US patent US5457956A1, 1995
[13] F. Formosa, "Nonlinear dynamics analysis of a membrane Stirling engine: Starting and stable operation", Journal of Sound and Vibration (2009) 326 (3-5), pp.794-808
[14] J. McEntee and L. Bowman, "Oscillating diaphragm, Technical proceedings of the 1999 Int. Conf. on Modeling and Simulation of Microsystems (1999), pp. 567-600
[15] M. Moran, D. Wesolek, B. Berhande, and K. Rebello, "Microsystem cooler developments," Tech. Rep. NASA/TM—2004-213307, NASA Glenn Research Center, 2004
[16] K. Jeong, "Free vibration of two identical circular plates coupled with bounded fluid", Journal of Sound and Vibration 260 (2003), pp. 653–670
[17] B. Desjardins, M.J. Esteban, C. Grandmont and P. Le Tallec, "Weak solutions for a fluid-elastic structure interaction model", Revista Mathematica complutense XIV 2 (2001), pp. 523-538
[18] F. Daneshmand, E. Ghavanloo, "Coupled free vibration analysis of a fluid-filled rectangular container with a sagged bottom membrane, Journal of Fluids and Structures 26 (2010), pp.236-252
[19] M. Chiba, H. Watanabe and H. F. Bauer, "Hydroelastic coupled vibrations in a cylindrical container with a membrane bottom containing liquid with surface tension", Journal of Sound and Vibration (2002) 251(4), pp. 717-740
[20] M. Haterbouch, R. Benamar, "The effects of large vibration amplitudes on the axisymmetric mode shapes and natural frequencies of clamped thin isotropic circular plates. Part I: iterative and explicit analytical solution for non-linear transverse vibrations", Journal of Sound and Vibration 265 (2003) 123–154.
[21] S. Timoshenko, "The theory of plates and shells", McGraw-Hill Classic Textbook Re-issue Series, 1959